\documentclass[twocolumn,superscriptaddress]{revtex4-2}

\usepackage{graphicx}
\usepackage{dcolumn}
\usepackage{bm}
\usepackage{amsmath}
\usepackage{float}
\usepackage{multirow,booktabs}
\usepackage[colorlinks=true,citecolor=blue,linkcolor=magenta]{hyperref}

\begin{document}
	
\title{Continuous and deterministic all-photonic cluster state of indistinguishable photons}
	
\author{Zu-En Su}
\author{Boaz Taitler}
\author{Ido Schwartz}
\author{Dan Cogan}
\author{Ismail Nassar}
\author{Oded Kenneth}
\author{Netanel H. Lindner}
\author{David Gershoni}
	
\affiliation{The Physics Department and the Solid State Institute, Technion-Israel Institute of Technology, Haifa 32000, Israel}
	
\date{\today}

\begin{abstract}
Cluster states are key resources for measurement-based quantum information processing. Photonic cluster and graph states, in particular, play indispensable roles in quantum network and quantum metrology. We demonstrate a semiconductor quantum dot based device in which the confined hole spin acts as a needle in a quantum knitting machine producing continuously and deterministically at sub-Gigahertz repetition rate single indistinguishable photons which are all polarization entangled to each other and to the spin in a one dimensional cluster state. By projecting two nonadjacent photons onto circular polarization bases we disentangle the spin from the photons emitted in between, thus continuously and deterministically preparing all-photonic cluster states for the first time.
We use polarization tomography on four sequentially detected photons to demonstrate and to directly quantify the robustness of the cluster's entanglement and the determinism in its photon generation. 
\end{abstract}

\maketitle

\section{Introduction}

Entanglement represents the essence of quantum mechanics and plays a key role in quantum information processing. Multipartite entanglement is required in many future applications such as quantum computation, quantum metrology and quantum communication. Such entanglement is usually quite fragile and may completely vanish even if only one particle interacts with the environment. A well-known class of quantum states, called cluster states \cite{briegel2001persistent}, exhibits high entanglement robustness. In these clusters some entanglement remains as long as more than half of the particles are not disentangled from the system. These cluster states are important resources for single-qubit-measurement-based quantum computing \cite{raussendorf2001one,raussendorf2007fault}, for memory-less quantum repeaters and for entanglement distribution between remote nodes \cite{zwerger2012measurement}. 

Cluster states have been demonstrated in various physical systems, including trapped ions \cite{mandel2003controlled,lanyon2013measurement}, cavity quantum electrodynamics \cite{thomas2022efficient}, continuous-variable modes \cite{yukawa2008experimental,armstrong2012programmable,yokoyama2013ultra,chen2014experimental,yoshikawa2016invited,cai2017multimode,larsen2019deterministic,asavanant2019generation}, and superconducting qubits \cite{wang201816,gong2019genuine} or related microwave photons \cite{besse2020realizing,ferreira2022deterministic}. 
Photonic cluster states, in particular, attract great attention since they provide precise single-qubit measurements, light-speed transmission and almost decoherence-free propagation. Few photons cluster states have been first probabilistically generated via the techniques of either parametric down-conversion \cite{walther2005experimental,zhang2006experimental,lu2007experimental} or four-wave mixing \cite{adcock2019programmable,vigliar2021error}.

As a promising platform for on-demand sources of both single photons \cite{dekel2000carrier} and entangled photon pairs \cite{akopian2006entangled}, semiconductor quantum dots (QDs) were proposed by Lindner and Rudolph \cite{lindner2009proposal} also as a venue for generation of linear photonic cluster states.
In their proposal the magnetic field induced precession of a QD confined electron's spin is used to entangle the polarizations of the sequentially emitted photons, which result from periodical optical excitation of the QD, timed with the spin precession. Schwartz \textit{et al.} \cite{schwartz2016deterministic} demonstrated the proposed scheme using resonant optical excitation of a precessing QD confined dark exciton (DE). The DE rather than the electron spin was chosen for this first proof of concept demonstration, mainly because the DE, as well as the heavy-hole (HH) spin's coherence time were found to be an order of magnitude longer than that of the electron spin \cite{cogan2018depolarization}.
More recently, Cogan \textit{et al.} \cite{cogan2023deterministic} replaced the entangling DE with a confined HH \cite{cogan2022spin} and demonstrated, for the first time, a cluster state composed of indistinguishable photons. Photon indistinguishability is indispensable for Bell-state measurements (BSMs), for measurement-based quantum communication \cite{azuma2015all,buterakos2017deterministic}, and for probabilistic scaling up of photonic clusters \cite{lindner2009proposal,cogan2023deterministic}. 

The demonstrations of Schwartz and Cogan \textit{et al.} \cite{schwartz2016deterministic,cogan2023deterministic} are based on the ability to deterministically control the QD confined spin. In principle, this approach allows for the generation of cluster states with an arbitrarily long sequence of photons, since  
as we unambiguously show below, the photons are deterministically generated at GHz repetition rate. The extraction and eventual detection rate of the emitted photons, however, is still about two orders of magnitude slower, due to the limited efficiency by which the light is extracted from the QD based device. The detection rates of events in which few sequential photons are detected, decreases exponentially with the number of photons, accordingly. 
Therefore, it is still quite challenging to demonstrate and use the produced cluster state. 

For demonstrating deterministic generation of the cluster state and quantifying its entanglement robustness, detection of events with three sequential photons is sufficient \cite{schwartz2016deterministic,cogan2023deterministic,toth2005entanglement}. 
However, since all-photonic clusters require disentanglement of the QD-spin qubit from the photonic qubits, 
demonstration of an all-photonic cluster requires detection of at least four sequential photons.  

Here, we use externally applied magnetic field in Voigt configuration to tune the entangling spin precession rate to achieve optimal entanglement length and to match exactly one quarter of the laser pulse repetition rate. As a result, the deterministically generated spin-photons-string is extremely long. 
The actual length of the all-photonic cluster state is then determined at will, simply by choosing the number of repeated excitations between the first and last photon which one detects after projecting their polarization on a circular polarization base. 
Then, by applying polarization tomography \cite{cogan2022pra} on the emitted photons in between, we demonstrate for the first time a deterministic all-photonic cluster state.  
At the same time the tomography also quantify the robustness of the entanglement between the cluster's photons. 
In addition, by performing polarization tomography between two next-nearest-neighbor photons, without detecting the nearest-neighbor photon in between, 
we demonstrate a novel way to quantify the determinism by which the entangled photons are generated. 

\section{Method}

\begin{figure*}[t]
	\includegraphics[width=2\columnwidth]{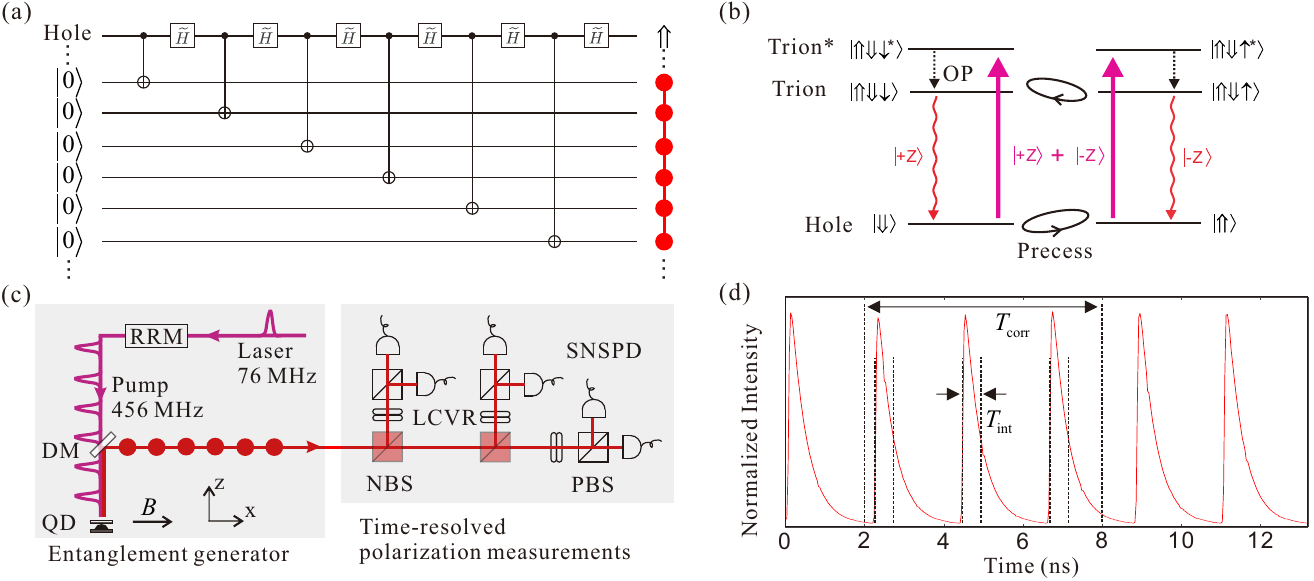}
	\caption{\label{fig:fig1}(a) A circuit diagram schematically describing quantum knitting machine as a periodic sequence of a two qubit $C_\text{not}$ gate and a single qubit $\widetilde{H}$ gate. The $C_\text{not}$ gate is realized by the optical pulsed excitation and subsequent emission, and the $\widetilde{H}$ gate by the timed spin precession. (b) Schematic description of the HH, Trion and excited Trion energy levels and the transitions between these levels, in the presence of external magnetic field in Voigt configuration (x-direction). The spin configuration of each level is presented where upward (downward) arrow represents (anti-) parallel spin direction relative to the optical direction $\left| Z \right\rangle $. Double (single) arrow represents HH (electron) spin.  Upward bold (downward wavy) arrows between the HH and Trion represent optical excitation (emission) while downward dotted arrows represent spin preserving optical phonon (OP) assisted relaxation of the excited Trion's electron. The exciting laser pulse is horizontally polarized in the direction of the field $\left| {X} \right\rangle $. The emitted photon polarization is entangled with the HH spin, thereby realizing the $C_\text{not}$ gate between the spin qubit and the emitted photon qubit.  (c) Schematic description of the experimental setup. The QD is located in a cryostat at $\sim$4 Kelvin subject to externally applied magnetic field and optically excited by resonant laser $\pi$-pulses. The repetition-rate multiplier (RRM) multiplies the laser's repetion rate from 76 MHz to 456 MHz, matching approximately four times the precession rate of the HH spin. The dichroic mirror (DM) spectrally separates the laser light from the emitted light, which is directed by two non-polarizing beams splitters (NBS) into three channels. Each channel contains two liquid crystal variable retarders (LCVRs), a polarizing beam splitter (PBS) and two superconducting nanowire single photon detectors (SNSPDs), enabling full polarization tomography of the detected photons. (d) Continuous time-resolved emission intensity as detected on each one of the six SNSPDs in c). }
\end{figure*}

The cluster state is generated by applying a periodic sequence of controlled-NOT ($C_\text{not}$) and rotated Hadamard (Eq.~\ref{eq:H}) gates on the HH spin ($\left|  \Uparrow  \right\rangle $), as shown in Fig.~\ref{fig:fig1}(a). 
The spin is confined in a single InAs/GaAs self-assembled QD which is well characterized by time-resolved polarization sensitive measurements \cite{cogan2022spin,cogan2020complete}. Fig.~\ref{fig:fig1}(b) schematically describes the generation of spin-photon entanglement. A $\pi$-pulse deterministically excites the HH to the positively-charged-trion ($\left| { \Uparrow  \Downarrow  \uparrow } \right\rangle $) with its electron in the first excited energy level. The electron decays to its respective ground level within about 10 ps by emitting a spin-preserving optical phonon \cite{benny2012excitation}. The trion then recombines radiatively within about 400 ps, leaving the HH at its ground level, while its spin is entangled with the polarization of the emitted photon. The emitted photons in this process are highly indistinguishable \cite{cogan2023deterministic} unlike the case in which the DE was used as entangler \cite{schwartz2016deterministic}. A dichroic mirror easily separates the emitted photons (966.45 nm) from the reflected resonant laser (945.99 nm) light, as shown in Fig.~\ref{fig:fig1}(c).

A weak in-plane magnetic field removes the Kramers' degeneracy between the two HH spin states, resulting in precessing of coherent superpositions of the HH spin eigenstates around the direction of the external field. The rotated Hadamard gate on the HH spin is realized by waiting one-quarter of its precession period, between two sequential excitation pulses: 
\begin{equation} \label{eq:H}
	\widetilde{H}(\left|  \Uparrow  \right\rangle) \longrightarrow \frac{\left(\left| \Uparrow \right\rangle+i \left| \Downarrow \right\rangle\right)}{\sqrt{2}}{\text{ ; }}\widetilde{H}(\left|  \Downarrow  \right\rangle )\longrightarrow \frac{\left(\left| \Downarrow\right\rangle+i\left| \Uparrow \right\rangle\right)}{\sqrt{2}}.
\end{equation}
As shown in Fig.~\ref{fig:fig1}(b), the selection rules of the optical transitions between two HH states and two positively-charged trion states are given by
\begin{equation} \label{eq:transition}
	\left|  \Uparrow  \right\rangle \overset {\left|  {-Z}  \right\rangle} \longleftrightarrow \left| { \Uparrow  \Downarrow  \uparrow } \right\rangle {\text{ ; }}\left|  \Downarrow  \right\rangle \overset {\left|  {Z}  \right\rangle} \longleftrightarrow \left| { \Uparrow  \Downarrow  \downarrow } \right\rangle,
\end{equation}
where $\left|  {\pm Z}  \right\rangle$ represent photons with right and left circular polarizations along the light propagation direction. A laser pulse with horizontal linear polarization $\left|  X  \right\rangle=(\left|  Z  \right\rangle+\left|  {-Z}  \right\rangle)/\sqrt{2}$ can coherently convert a coherent HH spin state $\alpha \left|  \Uparrow  \right\rangle + \beta \left|  \Downarrow  \right\rangle $  into a trion state with the same superposition amplitudes. The trion then radiatively decays into a spin-photon entangled state. Thus,  ideally, this action on the spin realizes a two qubits $C_\text{not}$ gate between the spin and the emitted photon:
\begin{equation} \label{eq:cnot}
	C_\text{not}( \alpha \left|  \Uparrow  \right\rangle + \beta \left|  \Downarrow  \right\rangle )  \longrightarrow \alpha \left|  \Uparrow  \right\rangle \left|  {-Z}  \right\rangle+ \beta \left|  \Downarrow  \right\rangle\left|  {Z}  \right\rangle .
\end{equation}
The process of realizing $\widetilde{H}$ and $C_\text{not}$ gate is repeated, in principle, indefinitely. This can be seen in the time resolved single-photon detection intensity resulting from this periodic continuous excitations presented in Fig.~\ref{fig:fig1}(d). 

In reality, however, the process deviates from the ideal description by two logical gates. There are two major contributions to this deviation: i) The finite radiative lifetime during which the HH in the ground state and the electron in the excited state continue to precess. ii) The decoherence in the electron and HH spin precession due to the hyperfine interaction with the nuclear spins in the QD vicinity. While the first imperfection increases with the magnetic field, which shortens the precession time relative to the radiative lifetime, the second decreases with the field. Therefore, these two imperfections result in limitation to the entanglement length of the cluster state, reaching optimal length of about 10 qubits at 0.09 Tesla. At this optimal field the excitation rate is approximately 456 MHz, which is also six times the repetition rate of the commercial pumped laser source \cite{cogan2023deterministic}. In Ref. \cite{cogan2022spin, cogan2023deterministic} we developed a realistic model which precisely describes the cluster generation.   

The emitted photons are divided by two non-polarizing beam-splitters into three channels as shown in Fig.~\ref{fig:fig1}(c). In each channel two liquid crystal variable retarders and a polarizing beam splitter (PBS) are used to select the polarization projection base.  Each output port of the PBS is equipped with a superconducting nanowire single photon detector (SNSPD) which records the detection time of the polarized photon either on the projection direction or on its complementary cross polarized direction. We note from Eq.~\ref{eq:transition} that a detection of a circularly polarized photon projects the state of the HH spin, thereby disentangles the spin qubit from the emitted photonic qubits.  

The detection times from the six SNSPDs relative to the periodic laser sync time are recorded and preliminary analyzed by a time-tagger. This six detectors setup is required for two reasons: i) It provides the means to record events in which multiple photons are sequentially detected, overcoming the deadtime of about 20 ns  of the SNSPDs, order of magnitude longer than the time between subsequent photon emissions (about 2 ns); ii) It allows full polarization tomography of three photons and partial tomography of even larger number of photons, as we present below.

For analyzing the experimental data, two temporal windows are defined, as can be seen in Fig.~\ref{fig:fig1}(d). i) The integration time ($T_\text{int}$) --- the time window following the excitation pulse in which a detected photon is associated with the exciting pulse. The larger the integration window is, the better is the photon detection efficiency.  However, due to the spin precession during the integration time, the larger the integration window is, the lower is the measured degree of polarization. The integration window that we choose for optimizing the analysis presented below (about 0.4 ns, - the trion's radiative lifetime) maximizes the product of the detection efficiency of two sequential photons and the measured degree of polarization of three sequentially detected photons  where the first and last photons are projected on a circular polarization base and the second photon on the rectilinear polarization base ($Z_{i-1}X_{i}Z_{i+1}$ the measured stabilizer - see below).

ii) The correlation time window ($T_\text{corr}$) --- the time between two detected photon events that the system registers and saves as a correlated two-photon event.  
The decision to store only events in which two sequential photons are detected, is required, since it reduces the amount of information that the system stores, by about 4 orders of magnitude than the output that the time tagging electronics produces.
By examining the stored times of correlated two-photon events, one can readily identify correlated three-photon events. 
More specifically, a three photon event occurs when two correlated two-photon events have one photon in common. 
Similarly, correlated four-photon events are also identified. 
We note here that since the system photon detection efficiency is about 1\%, the rate by which we detected correlated two-photon events amounts to about 50 kHz, and correlated three- and four-photon events to about 500 Hz and 5 Hz, respectively.
We choose $T_\text{corr}$ to include three sequential integration windows, thereby we store both nearest and next nearest neighbors in the emitted string of detected photons.  This in turn allows us to analyze multi sequential photon events, in which one photon between two detected photons was missing. 

\section{Results}

\begin{figure}[tb]
	\includegraphics[width=1\columnwidth]{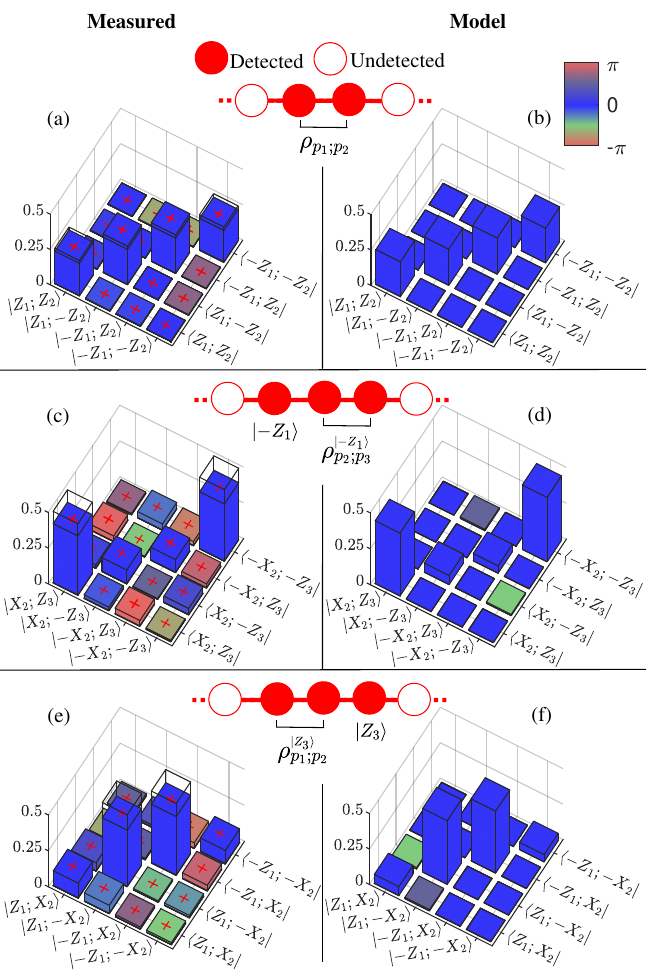}
	\caption{\label{fig:fig2} Measured [calculated] two-sequential photon polarization density matrices under various conditional circular polarization projections ($\left| { \pm Z} \right\rangle $) of neighboring photons. (a)[(b)] Two photon correlations without projection. Three sequential photon correlations, with earlier (c)[(d)] or later (e)[(f)] photon polarization projection. The height of the colored (uncolored) bars represent the absolute values of the measured (ideal) two-photon density matrix elements, while the color of the bars represent the phases of the density matrix elements. Error bars represent one standard deviations calculated from the propagated Poissonian statistics.}
\end{figure}

Before discussing the measurements, let us consider the expected results assuming an ideal system, in which the polarization projection is ideal and the logical $\widetilde{H}$ and $C_\text{not}$ gates are accurately defined by  Eq. \ref{eq:H} and Eq.~\ref{eq:cnot}, respectively.  
The measured polarization density matrix of every single photon $\rho _{p_1}$, where $p_i$ represents the $i^\text{th}$ sequentially detected photon by our system, is expected to be completely unpolarized  (see state No.~1 in Table ~\ref{tab:states}). This is because the confined hole spin is unpolarized to begin with and applying the logical gates on such a totally mixed spin state results in totally mixed spin-photon state. 
Likewise, the polarization density matrix of all the detected pairs of sequentially emitted photons, should also be totally mixed as described by the identity matrix $\rho _{p_1;p_2}$ of state No.~2 in Table~\ref{tab:states}. This totally mixed two-photon polarization state is indeed also evidenced by the measured [modelled] density matrix presented in Fig.~\ref{fig:fig2}(a)[(b)].
As expected, the negativity of the measured density matrix, which quantify the entaglement between the polarization of the two photons \cite{peres1996separability}, vanishes to within the experimental uncertainty.

We now turn to measure three-photon events where the first photon is projected on circularly polarized basis and the last two photons are measured on all bases.
Ideally, the detection of the first photon on left hand circular polarization $ \left| {{-Z}_1} \right\rangle $ heralds the confined spin on $\left|  \Uparrow  \right\rangle $ state, according to Eq.~\ref{eq:transition}.
The application of the $\widetilde{H}$, $C_\text{not}$, $\widetilde{H}$ and again $C_\text{not}$ gates on the spin is described as follows:
\begin{equation} \label{eq:spp}
	C_\text{not} \cdot \widetilde{H} \cdot C_\text{not} \cdot \widetilde{H} ( \left|  \Uparrow  \right\rangle) \longrightarrow\psi _{p_2;p_3;s}^{\scriptscriptstyle{\left|  {-Z}_1 \right\rangle}}
\end{equation}
where the resulting spin-two-photon state  $ \psi _{p_2;p_3;s}^{\scriptscriptstyle{\left| {-Z}_1 \right\rangle}} $ is given in state No.~3 in Table.~\ref{tab:states}.
In calculating state No. 3, we expressed the rectilinear polarization base in terms of the circular polarization base:
$\left| X \right\rangle =\frac {1}{\sqrt{2}} (\left| Z \right\rangle +\left| {-Z} \right\rangle)$; $\left| {-X} \right\rangle = \frac {i}{\sqrt{2}}(\left| Z \right\rangle -\left| {-Z} \right\rangle).$
The polarization density matrix between the second and third photon $\rho^{\scriptscriptstyle {\left| {-Z}_1 \right\rangle}}_{p_2;p_3}$ is described by state No.~4 in Table~\ref{tab:states}, simply by tracing out the spin state. The obtained density matrix in this case is polarized, but the two photons are not expected to be entangled. The actually measured and modelled matrices are shown in Fig.~\ref{fig:fig2}(c) and (d), respectively. They are similar but deviate from the ideally expected one from Eq.~\ref{eq:spp}. The fidelity between the measured density matrix and the ideal one and the modelled one \cite{cogan2023deterministic} is $0.81 \pm 0.01$ and $0.98 \pm 0.01$, respectively. Indeed, the negativity of the measured density matrix vanishes to within the experimental uncertainty. This is expected, since the state of the entangled spin is traced out.

It is worth noting that by applying time reversal symmetry arguments, one can in a similar way, consider the quantum state of the first and second photons when the last photon is detected in circular polarization. It follows that in this case the resulting two-photon polarization density matrix  $\rho _{p_1;p_2}^{\scriptscriptstyle {\left| Z_3 \right\rangle}}$ is given by state No.~5 in Table~\ref{tab:states}. The measured [modeled] polarization density matrix are presented in Fig.~\ref{fig:fig2}(e)[(f)]. The fidelity of the measured density matrix to the ideal and modelled ones is $0.80 \pm 0.01$ and $0.98 \pm 0.01$, respectively. Here as well, the two photons are polarized, but not entangled and the negativity of the polarization density matrix vanishes.

\begin{figure}[tb]
	\includegraphics[width=1\columnwidth]{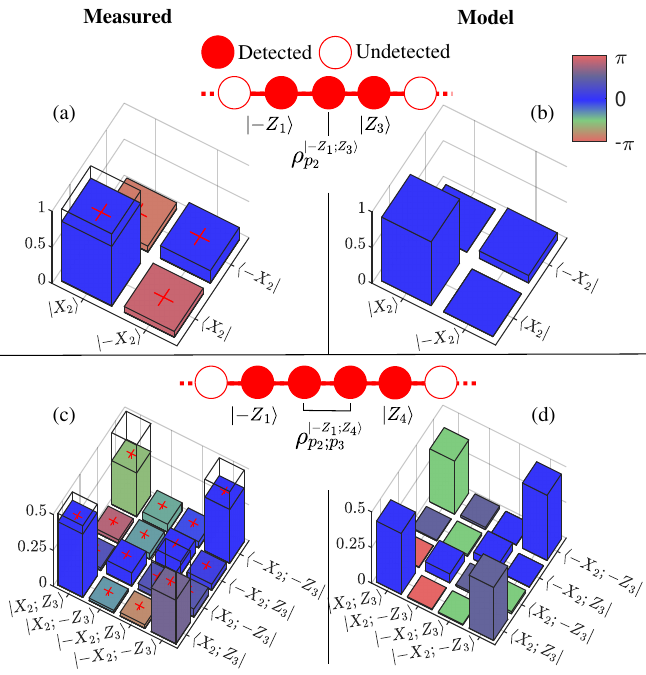}
	\caption{\label{fig:fig3} Measured [modelled] polarization density matrices conditioned on detecting circularly polarized photons before and after the measured photons. (a)[(b)] Polarization density matrix of a single photon. (c)[(d)] Polarization density matrix of two sequential photons. The height of the colored (uncolored) bar represents the absolute value of the measured (ideal) density matrix element, while the color of the bar represents the phase of the density matrix element. Error bars represent standard deviation of the experimental uncertainty calculated assuming Poissonian photon statistics.}
\end{figure}

Linear cluster states are stabilized by a series of three-qubit operators $Z_{i-1}X_{i}Z_{i+1}$ \cite{toth2005entanglement}. Measurement of the stabilizer is used to characterize the robustness of the entanglement in the cluster \cite{cogan2023deterministic}. The stabilizer measurement consists of two $Z$ measurements that effectively isolate the middle qubit from the rest of the chain.
For an ideal cluster one should get an expectation value of 1 for the stabilizer. To demonstrate it we proceed by considering the spin-two-photon state $\psi _{p_2;p_3;s}^{\scriptscriptstyle{\left|  {{-Z}_1}  \right\rangle}} $ from Eq.~\ref{eq:spp} (state No.~3 in Table~\ref{tab:states}). By projecting the third photon on a right hand circular polarization ($\left| {Z_3}\right\rangle $) one selects the 
part with the $ \left| \Downarrow \right\rangle $ spin state of the entangled state $\psi _{p_2;p_3;s}^{\scriptscriptstyle{\left|  {{-Z}_1}  \right\rangle}} $ and leaves the second, disentangled photon, in a pure polarization state:  $\psi _{p_2}^{\scriptscriptstyle {\left| {-Z}_1;Z_3 \right\rangle}}= \left|  X_2 \right\rangle $. Thus, the disentangled photon is expected to be fully rectilinearly polarized (state No.~5 in Table~\ref{tab:states}). In Fig.~\ref{fig:fig3}(a)[(b)] we present the measured [modelled] results. 
The degree of measured [modelled] rectilinear polarization of the second photon is given by 
\begin{equation}
	P_{p_2}^{\scriptscriptstyle {\left| {-Z}_1;Z_3\right\rangle}}=\frac{C_{\left|{{X}_2}\right\rangle}-C_{\left|{{-X}_2}\right\rangle}}{C_{\left|{{X}_2}\right\rangle}+C_{\left|{{-X}_2}\right\rangle}}
	=0.65 \pm 0.01 [0.79 \pm 0.01],
\end{equation}
where $C_{\left|{\pm{X}_2}\right\rangle}$ refer to the number of counts per unit time, projected on rectilinear polarizations $\left|{\pm{X}_2}\right\rangle$.

\begin{table*}[!htb]
	\centering
	\caption{Ideal spin-photons quantum states and polarization density matrices.}
	\label{tab:states}
	\begin{ruledtabular}
		\begin{tabular}{l l l}
			No. & States & Expressions\footnotemark[1] \\
			\midrule
			1 & $\rho _{p_1} $ & $\frac{1}{2}(\left| {Z_1} \right\rangle \left\langle {Z_1} \right| + \left| {{-Z}_1} \right\rangle \left\langle {{-Z}_1} \right|)$ \\
			2 & $\rho _{{p_1};{p_2}} $ & $ \frac{1}{4}(\left| {{Z}_1};{{Z}_2} \right\rangle \left\langle {{Z}_1};{{Z}_2} \right| + \left| {{Z}_1};{{-Z}_2} \right\rangle \left\langle {{Z}_1};{{-Z}_2} \right|+\left| {{-Z}_1};{{Z}_2} \right\rangle \left\langle {{-Z}_1};{{Z}_2}\right| + \left| {{-Z}_1};{{-Z}_2} \right\rangle \left\langle {-Z_1};{-Z_2} \right|)$  \\
			\midrule
			3 & $ \psi _{{p_2};{p_3};{s}}^{\scriptscriptstyle{\left| {-Z}_1 \right\rangle}} $ & $\frac{1} {\sqrt{2}} \left( \left|{{-X}_2};{{-Z}_3} \right\rangle \left|  \Uparrow  \right\rangle + \left| {{X}_2};{{Z}_3} \right\rangle\left| \Downarrow \right\rangle \right) $ 	\\
			4 & $ \rho _{{p_2};{p_3}}^{\scriptscriptstyle{\left| {-Z}_1 \right\rangle}} $ & $\frac{1}{2}(\left|{{-X}_2};{{-Z}_3}\right\rangle\left\langle{-X}_2;{-Z}_3\right| + \left|{X}_2;{Z}_3\right\rangle\left\langle {X}_2;{Z}_3\right|)$    \\
			5 & $\rho _{{p_1};{p_2}}^{\scriptscriptstyle{\left| {Z}_3 \right\rangle}}$ & $\frac{1}{2}(\left|{{Z}_1};{{-X}_2}\right\rangle\left\langle{{Z}_1};{{-X}_2}\right|+\left|{{-Z}_1};{{X}_2}\right\rangle\left\langle{{-Z}_1};{{X}_2}\right|)$  \\
			\midrule
			6 & $\psi _{p_2}^{\scriptscriptstyle{\left| {-Z}_1;{Z}_3 \right\rangle}}$ & $\left|  X_2 \right\rangle $    \\
			7 & $\psi _{{p_2};{p_3};{p_4};{s}}^{\scriptscriptstyle{\left|  {-Z}_1 \right\rangle}} $ & $ \frac {1}{2} \left[\left (i\left| {-X}_2;{-Z}_3 \right\rangle -\left|{X}_2;{Z}_3 \right\rangle \right) \left|{-Z}_4\right\rangle \left|\Uparrow\right\rangle + \left (-\left|{-X}_2;{-Z}_3 \right\rangle +i\left|{X}_2;{Z}_3 \right\rangle \right) \left|{Z}_4 \right\rangle \left| \Downarrow \right\rangle\right] $ \\
			8 & $\psi _{{p_2};{p_3}}^{\scriptscriptstyle{\left|  {-Z}_1;{Z}_4 \right\rangle}}$ & $ \frac {1}{\sqrt{2}}\left( -\left|{{-X}_2};{{-Z}_3} \right\rangle +i\left|{{X}_2};{{Z}_3} \right\rangle\right)$   \\
			\midrule
			9 & $\psi _{p_3;s}^{\scriptscriptstyle{\left| {-Z}_1 \right\rangle}}$\footnotemark[2] & $\left|  {Z}_3 \right\rangle \left| \Downarrow \right\rangle$   \\
			10 & $\rho _{p_3}^{\scriptscriptstyle{\left| {-Z}_1 \right\rangle}}$ &  $\frac{1}{2}(\left|{{Z}_3}\right\rangle\left\langle{{Z}_3}\right|+\left|{{-Z}_3}\right\rangle\left\langle{{-Z}_3}\right|)$  \\
			11 & $\psi _{{p_2};{p_4};{p_5};s}^{\scriptscriptstyle{\left|  {-Z}_1 \right\rangle}}$\footnotemark[2] & $ \frac {1}{2} \left[-\left (\left| {X}_2;{-X}_4 \right\rangle +\left|{-X}_2;{X}_4 \right\rangle \right) \left|{-Z}_5\right\rangle \left|\Uparrow\right\rangle + \left (\left|{X}_2;{-X}_4 \right\rangle -\left|{-X}_2;{X}_4 \right\rangle \right) \left|{Z}_5 \right\rangle \left| \Downarrow \right\rangle\right] $ \\
			12 & $\psi _{{p_2};{p_4}}^{\scriptscriptstyle{\left|  {-Z}_1;{Z}_5 \right\rangle}}$\footnotemark[2] & $ \frac {1}{\sqrt{2}}\left( \left|{X}_2;{-X}_4 \right\rangle -\left|{-X}_2;{X}_4 \right\rangle\right)$  \\
			13 & $\psi _{{p_2};{p_3};{p_4};{p_5};{s}}^{\scriptscriptstyle{\left|  {-Z}_1 \right\rangle}} $ & $ \frac {1}{2} \left[-\left (\left| {X}_2;{Z}_3;{X}_4 \right\rangle +\left|{-X}_2;{-Z}_3;{-X}_4 \right\rangle \right) \left|{-Z}_5\right\rangle \left|\Uparrow\right\rangle + \left (\left|{X}_2;{Z}_3;{-X}_4 \right\rangle -\left|{-X}_2;{-Z}_3;{X}_4 \right\rangle \right) \left|{Z}_5 \right\rangle \left| \Downarrow \right\rangle\right] $ \\
			14 & $\rho _{{p_2};{p_4}}^{\scriptscriptstyle{\left|  {-Z}_1;{Z}_5 \right\rangle}}$ & $\frac{1}{2}(\left|{{X}_2};{{-X}_4}\right\rangle\left\langle{{X}_2};{{-X}_4}\right|+\left|{{-X}_2};{{X}_4}\right\rangle\left\langle{{-X}_2};{{X}_4}\right|)$ \\
		\end{tabular}
	\end{ruledtabular}    
	\footnotetext[1]{$ \left| \Uparrow (\Downarrow) \right\rangle$ is the spin state and $\left| P_i  \right\rangle$ is the polarization state of the $i^{th}$- sequentially detected photon.}
	\footnotetext[2]{Only these three states refer to the case of $D=0$ while all others refer to the case of $D=1$.}
\end{table*} 

Let us generalize the case to four sequential photons. In this case the middle two photons are disentangled from the spin on both ends and therefore the two photons are expected to be entangled. This can be understood by applying the protocol once more on $\psi _{p_2;p_3;s}^{\scriptscriptstyle{\left| {-Z}_1 \right\rangle}} $:
\begin{equation}
	C_\text{not} \cdot \widetilde{H} (\psi _{p_2;p_3;s}^{\scriptscriptstyle{\left| {-Z}_1 \right\rangle}}) \longrightarrow\psi _{p_2;p_3;p_4;s}^{\scriptscriptstyle{\left|  {-Z}_1 \right\rangle}},
\end{equation}
where the final state is given in state No.~7 in Table~\ref{tab:states}. Now, by detecting the fourth photon in right hand circular polarization the hole spin is disentangled and the second and third photon state $ \psi _{p_2;p_3}^{\scriptscriptstyle{\left|  {-Z}_1;Z_4 \right\rangle}}$  is given by state No.~8 in Table~\ref{tab:states}. 
The corresponding measured polarization density matrix and our realistic model calculations are shown in Fig.~\ref{fig:fig3}(c) and (d), respectively. Like in Fig.~\ref{fig:fig2}(c)-(d), the two photons are polarized either $\left|{{X}_2};{{-Z}_3} \right\rangle$ or $\left|{{-X}_2};{{Z}_3} \right\rangle$ but the ``phase'' between these two possibilities is well defined. This is evidenced by the appearance of the non-diagonal matrix elements, with the expected phase in the measured and modelled density matrices. The negativity of the measured (modelled) density matrix amounts to $0.22\pm0.01$ ($0.32\pm0.01$), and the fidelity between the measured and the ideal (modelled) density matrices is $0.71\pm0.01$ ($0.94\pm0.02$).

\section{Discussion}

The measured results present, in fact, for the first time a realization of an all-photonic cluster state. Though the cluster state that we demonstrate is very short, (only two entangled photons), yet, the concept is clearly demonstrated. Since as we demonstrate below, the photon generation in our device is deterministic, the only remaining technological challenge, which prevent us from using our all-photonic cluster state for applications is to deterministically detect the emitted photons. The efficiency by which the emitted photons are collected from our planar microcavity embedded QD is more than 20\%. We lose about additional order of magnitude in our experimental setup before we detect the collected photons. With this efficiency, we detect correlated four-photon events in a few Hertz rate, thereby limiting our ability to experimentally demonstrate longer photonic cluster states. Notable improvements of light harvesting efficiencies have been recently reported \cite{senellart2017high,liu2018high,tomm2021bright}. Successful realization of these improvements will enable demonstration of much longer photonic cluster states and their use in quantum communication protocols \cite{azuma2015all,buterakos2017deterministic}.  

\begin{figure}[tb]
	\includegraphics[width=1\columnwidth]{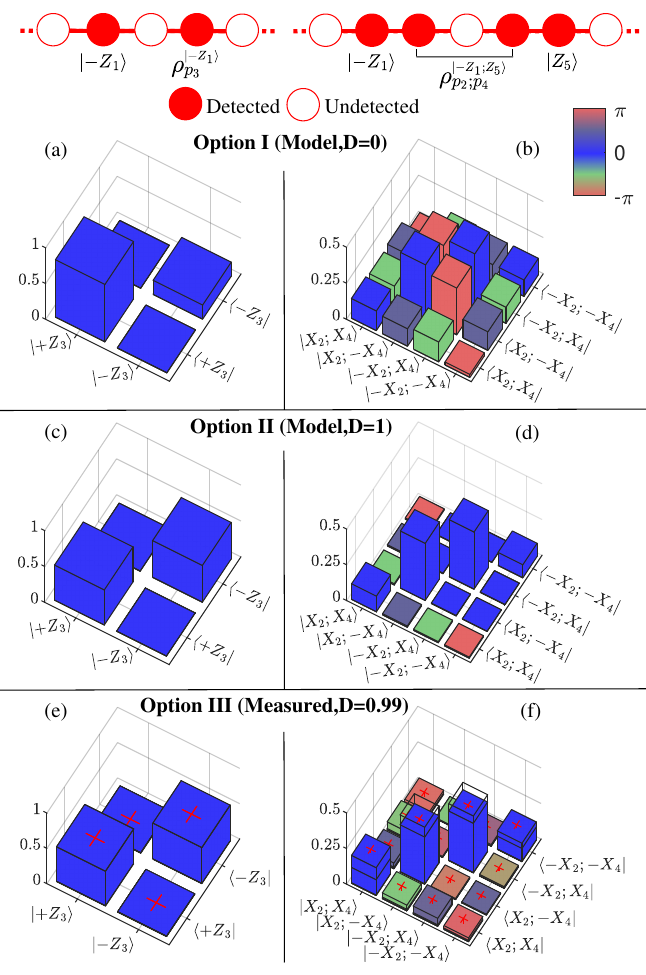}
	\caption{\label{fig:fig4} Measured and modelled polarization density matrices for events in which one intermediate sequential photon was not detected while the first and last photon are detected with circular polarization $\left| { \pm Z} \right\rangle $. (a) and (b) are modelled, assuming that the spin was not excited ($D=0$). (c) and (d) are modelled assuming deterministic generation ($D=1$). Colored bars in (e) and (f) represent the measured results, while uncolored bars represent the best fitted model calculations with $D$ as fitting parameter. In the fitting the modelled matrices with $D=0$ are weighted by ($1-D$) and those with $D=1$ are weighted by $D$. The color of a bar represents its phase.}
\end{figure}

The efficiency of the emitted light detection can be quite accurately measured using well calibrated light sources. The determinism by which the pulsed excitation of the QD results in actual photon absorption and spin excitation is more difficult to measure. A common practice is to use the measured efficiency of the detection and then to experimentally verify that the exciting pulse intensity produces a $\pi$-pulse. Under these conditions, from the ratio between the excitation rate and the detection rate and by comparison with the detection efficiency one can estimate the determinism of the excitation. 

Here we present another novel method to measure the photon excitation efficiency, independently from the photon detection efficiency. 
We do that, as explained above,  by setting the correlation window $T_\text{corr}$ wide enough to include three sequential excitations of the QD. This allow us to record correlation events in which one photon is not detected while photons before and after are detected. 

Conceptually, the absence of a detected photon can be attributed either to \textbf{i)} non-deterministic excitation, or to \textbf{ii)} non-deterministic detection, or to \textbf{iii)} both.
The three cases can be quite accurately quantified using the consideration below.
The detection of $\left|  {-Z}_1  \right\rangle$ polarized photon heralds the spin in the $\left|  \Uparrow  \right\rangle$ state. 
The spin then precesses for a quarter of a precession period until the second excitation pulse is applied. 
Let us consider \textbf{Option i} first: while the second excitation pulse fails to excite the spin (the determinism $D=0$), the spin continues to coherently precess, (since the excitation rate is at least an order of magnitude shorter than the spin's coherence time \cite{cogan2018depolarization, cogan2022spin}) for another quarter of a precession until the next excitation pulse is applied. Under these circumstances the single spin qubit gate $\widetilde{H}$ becomes  $\widetilde{H}\cdot \widetilde{H}=\widetilde{H}^2$ where its effect on the spin is given by:
\begin{equation} \label{eq:HH}
	\widetilde{H}^2(\left|  \Uparrow  \right\rangle) \longrightarrow i\left| \Downarrow\right\rangle {\text{ ; }}\widetilde{H}^2(\left|  \Downarrow  \right\rangle )\longrightarrow i\left| \Uparrow\right\rangle.
\end{equation}
The excitation of the spin by the third pulse then results in application of the $C_\text{not}$ gate (Eq.~\ref{eq:cnot}) which leads to the spin-photon product state
$\psi _{p_3;s}^{\scriptscriptstyle{\left| {-Z}_1 \right\rangle}} = \left | {Z}_3 \right\rangle\left | \Downarrow \right\rangle $ (state No.~9 in Table~\ref{tab:states}). The emitted photon must therefore be right-hand circularly polarized ($\psi _{p_3}^{\scriptscriptstyle{\left| {-Z}_1 \right\rangle}} = \left|  {Z}_3 \right\rangle $). 
Fig.~\ref{fig:fig4}(a) presents the modelled polarization density matrix of the photon for this case, presenting indeed highly right-hand circularly polarized photon.
In \textbf{Option ii} the excitation pulse sucessfully excites the spin ($D = 1$) but the emitted photon is simply undetected, and has to be traced out. The emitted photon resulting from the third excitation in this case is completely unpolarized given by the density matrix No.~10 in Table~\ref{tab:states}. 
The modelled density matrix is presented in Fig.~\ref{fig:fig4}(c). 
In \textbf{Option iii} the determinism factor is $0<D<1$, and the density matrix of the emitted photon is given by: 
\begin{equation}
	\rho^{\text{(iii)}}=(1-D)\rho^{\text{(i)}} +D\rho^{\text{(ii)}} 
\end{equation}
where $\rho^{\text{(i)}}$, and $\rho^{\text{(ii)}}$ are the modelled density matrices in Option i and ii, respectively. The experimentally measured density matrix depicted in Fig.~\ref{fig:fig4}(e) closely resembles the modeled result of Option ii shown in Fig.~\ref{fig:fig4}(c). The two density matrices exhibit a fidelity of $0.99\pm0.01$, a value that can be further enhanced by adopting the model of Option iii, which incorporates a determinism factor of $D=0.99\pm 0.01$. 

To further validate the above method, we conducted a five-pulse experiment, as outlined in the upper-right corner of Fig.~\ref{fig:fig4}. In this scenario, we disentangle the spin of both sides of the photon strings by detecting the first and fifth photons projected on circular polarization bases. Subsequently, we measure the polarization density matrix of the second and fourth photons while leaving the third photon undetected. Similar to the three-pulse experiment, we initialize the spin in a $\left|  \Uparrow  \right\rangle$ state and consider three different options. In \textbf{Option i} the third pulse does not excite the spin ($D=0$), and it continues to precess coherently. The final spin-three-photons state resulting from applying the following sequence of gates on the spin: 
\begin{equation}
	C_\text{not} \cdot \widetilde{H} \cdot C_\text{not} \cdot \widetilde{H}^2 \cdot C_\text{not} \cdot \widetilde{H} ( \left|  \Uparrow  \right\rangle) \longrightarrow\psi _{p_2;p_4;p_5;s}^{\scriptscriptstyle{\left|  {-Z}_1 \right\rangle}}
\end{equation}
is given by state No.~11 in Table~\ref{tab:states}.
Projecting the fifth photon of state No.~11 on $\left|  Z \right\rangle$ polarization selects the part of the wavefunction with $\left|  \Downarrow  \right\rangle$ spin, leaving the second and fourth emitted photons in a maximally entangled state, as described by state No.~12 in Table~\ref{tab:states}. The corresponding, modelled density matrix, is shown in Fig.~\ref{fig:fig4}(b). 
In \textbf{Option ii} the third pulse successfully excites the spin ($D=1$), and the resulting spin-four-photon state
is given by state No.~13 in Table~\ref{tab:states}. After projecting the fifth photon on $\left|  Z \right\rangle$ circular polarization and tracing out the missing third photon, the second and fourth photons are left in a classically correlated state described by the density matrix No.~14 in Table~\ref{tab:states}. The modelled density matrix for this case is shown in Fig.~\ref{fig:fig4}(d). 
In \textbf{Option iii} the third pulse excites the spin in a probabilistic way ($0<D<1$). The measured density matrix, shown in Fig.~\ref{fig:fig4}(f), has a fidelity of $0.68\pm 0.02$ with the ideal one (state No.~14 in Table~\ref{tab:states}). By optimizing the weighted value $D$ of the density matrix for Options i and ii, we achieve the highest fidelity between the measured density matrix and the modelled one in Option iii, reaching a value of $0.97\pm0.01$ when $D$ equals $0.99\pm 0.16$. It can be seen that the three-pulse and the five-pulse experiments share a similar determinism factor. The larger error in the five pulse experiment is mainly due to the reduced statistics of four- rather than three-correlated photon detected events.  We note also that the negativity in the measured two-photons polarization density matrix vanishes to within the experimental uncertainty. This is expected for Option ii but not for Option i. 

The measured degree of rectilinear polarization of the second and fourth photons is given by $P_{{p_2};{p_4}}^{\scriptscriptstyle{\left|  {-Z}_1;{Z}_5 \right\rangle}}=-0.35 \pm 0.03$, defined as:
\begin{equation}
	\frac{C_{\left|{{X}_2;{X}_4}\right\rangle}-C_{\left|{{X}_2;{-X}_4}\right\rangle}-C_{\left|{{-X}_2;{X}_4}\right\rangle}+C_{\left|{{-X}_2;{-X}_4}\right\rangle}}{C_{\left|{{X}_2;{X}_4}\right\rangle}+C_{\left|{{X}_2;{-X}_4}\right\rangle}+C_{\left|{{-X}_2;{X}_4}\right\rangle}+C_{\left|{{-X}_2;{-X}_4}\right\rangle}},
\end{equation}
where $C_{\left|{\pm{X}_2;\pm{X}_4}\right\rangle}$ refer to the number of coincidence counts per unit time projected on rectilinear polarization $\left|{\pm{X}_2;\pm{X}_4}\right\rangle$.

Finally it is interesting to note that the five-pulse experiment provides a a five-qubit stabilizer:
\begin{equation}
	{Z_1}{X_2}{I_3}{X_4}{Z_5}=Z_1X_2Z_3\cdot Z_3X_4Z_5=|Z_1X_2Z_3|^2,
\end{equation}
where $I_3=Z_3\cdot Z_3$ is the identity operator applied to the undetected third photon. 
Therefore it is not surprizing that  the value of $\left | P_{{p_2};{p_4}}^{\scriptscriptstyle{\left|  {-Z}_1;{Z}_5 \right\rangle}} \right |$ is approximately given by the value of $\left | P_{p_2}^{\scriptscriptstyle {\left| {-Z}_1;Z_3\right\rangle}} \right |^2$.
It means that if the generation is indeed deterministic, in principle it is possible to loose one out of five photons and still quantify the expectation value of  the three-qubit stabilizer. This is not possible, for example, if the generation is probabilistic like the case of spontaneous parametric down-conversion and entangling gate based on post-selection \cite{walther2005experimental,zhang2006experimental,lu2007experimental,adcock2019programmable,vigliar2021error,istrati2020sequential,li2020multiphoton}. 

\section{Conclusion}

In summary, we demonstrate for the first time  continuous and deterministic generation of an all photonic cluster state of indistinguishable photons. 
The cluster state is generated by periodic excitation of a semiconductor quantum dot confined hole spin preccesing in an externally applied magnetic field in Voigt configuration. The spin is excited by rectilinearly polarized resonant optical $\pi-$ pulses at a sub-Gigahertz rate. The spin is disentangled from the photonic state by polarization projection of the first and last photon of an emitted photon string on a circular polarization base. The demonstration of the photon entanglement and the determinism of their generation are based on polarization tomography measurements and analysis of two photon density matrices of up to four sequentially detected photon events. 
Feasible future improvements of the device's light harvesting efficiency will make the device useful for measurement-based quantum information processing.

~\\
{\em Acknowledgments---}   This research was supported by the Israeli Science Foundation (ISF - grant No. 1933/23)
the European Research Council (ERC-Grant No. 695188) and the German Israeli Research Cooperation (DIP - grant No. DFG-FI947-6-1). 

\bibliography{continuous}

\end{document}